\documentclass[12pt]{article}

\newcommand{\beq}{\begin{equation}}
\newcommand{\eeq}{\end{equation}}
\newcommand{\beqa}{\begin{eqnarray}}
\newcommand{\eeqa}{\end{eqnarray}}

\def\ie{{\it i.e.}}

\def\gsim{\ \rlap{\raise 2pt \hbox{$>$}}{\lower 2pt \hbox{$\sim$}}\ }
\def\lsim{\ \rlap{\raise 2pt \hbox{$<$}}{\lower 2pt \hbox{$\sim$}}\ }
\usepackage{feynmp}                       

\usepackage{epsfig}

\usepackage[dvips]{color}                 
\usepackage{graphicx}                     
\usepackage{amssymb}

 \newcommand{\non}{\nonumber}

\newcommand{\ii}{\mathrm{i}} 
 \newcommand{\T}{\mathrm{T}}

 \newcommand{\pv}{\vec{\,\!p}\!\:{}}

\newcommand{\MeV}{\mathrm{MeV}} \newcommand{\fm}{\mathrm{fm}}

  \newcommand{\de}{\partial} \newcommand{\dev}{\vec{\de}}

\begin{document}
\begin{fmffile}{higfeyn}
  \fmfset{curly_len}{2mm} \fmfset{dash_len}{1.5mm} \fmfset{wiggly_len}{3mm}
  \newcommand{\feynbox}[2]{\mbox{\parbox{#1}{#2}}}
\newcommand{\fs}{\scriptstyle} 
\newcommand{\hq}{\hspace{0.5em}} \newcommand{\hqm}{\hspace{-0.25em}}

\fmfcmd{vardef ellipseraw (expr p, ang) = save radx; numeric radx; radx=6/10
  length p; save rady; numeric rady; rady=3/10 length p; pair center;
  center:=point 1/2 length(p) of p; save t; transform t; t:=identity xscaled
  (2*r*h) yscaled (2*rady*h) rotated (ang + angle direction length(p)/2 of
  p) shifted center; fullcircle transformed t enddef;
  style_def ellipse expr p= shadedraw ellipseraw (p,0); enddef;}
  
%


\sloppy
\hfill{\phantom{.}}

\rightline{ISU-HET-01-4}
\rightline{nucl-th/0104088}
\rightline{April, 2001}

\vspace{1.0cm}

\begin{center}
{\Large {\bf Effective Range in Doublet S-wave\\
\vskip 0.3truecm
 Neutron-Deuteron Scattering}}

\vspace{1.0cm}

{\large F. Gabbiani}
\bigskip

{\it Department of Physics and Astronomy, Iowa State University, Ames, IA 50014, USA\\}

\medskip

{Email: \tt fg@eft.physics.iastate.edu}
\end{center}

\vspace{1cm}

\begin{abstract}

The effective field theory approach is applied to the three-nucleon
process of $S=1/2$ neutron-deuteron scattering in the S-wave,
including the effective range parameters summed at all orders. This is
achieved through a modification of the power counting, following a
recent suggestion in the literature. It is shown that with this
procedure, while the convergence of the loop integrals improves, one
cannot meaningfully include a three-body effective term to describe
low-energy data affected by the presence of the triton bound state.
\end{abstract}

\vspace{0.5cm}

\section{Introduction}
Recently, three-body interactions in nucleon systems have attracted
considerable attention, both within more classical approaches based on potential
models \cite{hub} and in the context of effective field theories \cite{weinberg,KSW}
(EFT's) \cite{bosons,3stooges,hm,gegelia,3stooges2}.

The three-nucleon system is a natural test-ground for the
understanding of nuclear forces that has been reached in the
two-nucleon system. A convenient example is given by neutron-deuteron
scattering \cite{3stooges,3stooges2,pbhg}, since in this case Coulomb
interactions are negligible. Neutron-deuteron scattering involves two
$S$-wave channels, corresponding to total spin $S=3/2$ and $S=1/2$. In
the $S=3/2$ channel all spins are aligned and the two-nucleon
interactions are only in the $^3 \mathrm{S}_1$ partial wave. The
two-body interaction is attractive but the Pauli exclusion principle
forbids the three nucleons to be at the same point in
space. Therefore this channel is insensitive to short-distance
physics and disallows a three-body bound state. Thus
one is able to obtain precise ($\sim 4\%$) predictions in a
straightforward way \cite{3stooges,3stooges2,pbhg}. In the $S=1/2$
channel two-nucleon interaction can take place either in the $^3
\mathrm{S}_1$ or in the $^1 \mathrm{S}_0$ partial waves. This leads to
an attractive interaction which implies a three-body bound state, the
triton. The $S=1/2$ channel also shows a strong sensitivity to
short-distance physics as the Pauli principle does not apply.

In this note EFT is applied to the latter channel, but including the
effective range parameters for the $^3 \mathrm{S}_1$ and $^1
\mathrm{S}_0$ $NN$ scattering partial waves in
the calculation, following a suggestion proposed in \cite{besa}. In
that paper the authors argued that taking both the scattering length
$a$ and the effective range $r$ in the $NN$ scattering effective range
expansion of order $Q^{-1}$, and summing range corrections to all
orders, improves convergence and may solve problems encountered in the
pionful theory \cite{FMS} with KSW power counting \cite{KSW}. The
authors then successfully applied this modified power counting to a
few two-body processes. It is therefore important to test this new
scheme on a three-body system. It will be shown that the new power
counting improves the convergence of the loop integrals necessary for
the $nd$ scattering calculation, so that the cutoff introduced in
\cite{bosons,3stooges2} is no longer needed for the system of integral
equations {\it without} a three-body interaction term. However, even
this improved ultraviolet behavior of the kernels in the integral
equations is not enough, since adding a three-body interaction
introduces new divergences which make the equations insensitive to the
three-body term in the Lagrangian, as it will be shown below.

This independence of the amplitudes with respect to an additional
parameter prevents us from applying the fitting procedure of
ref.~\cite{3stooges} and therefore from predicting the energy
dependence of the $nd$ phase shifts.

\section{Formalism}

It is convenient to use a Lagrangian \cite{david,pbhg}
expressed in terms of two auxiliary fields $d^i$ and $t^A$ with the quantum
numbers of the deuteron and of a dibaryon field in the ${}^1\mathrm{S}_0$
channel of $NN$ scattering respectively:
\begin{eqnarray}\label{dlag}
   \mathcal{L}_{Nd}&=&N^\dagger (\ii \partial_0
     +\frac{\dev^2}{2 M})N+\non\\
   &&+d^{i  \dagger} \left[-(\ii \partial_0
     +\frac{\nabla^2}{4
  M})-\Delta^{(-1)}_d-\Delta^{(0)}_d\right]d^i
   +\;y_d\left[d^{i \dagger} (N^\T P^i_d
     N) +\mathrm{h.c.}\right] +\\
      &&+t^{A \dagger} \left[-(\ii \partial_0 +
        \frac{\nabla^2}{4
  M})-\Delta^{(-1)}_t-\Delta^{(0)}_t\right]t^A
   +\;y_t\left[t^{A \dagger} (N^\T P^A_t
     N) +\mathrm{h.c.}\right] +
   \dots\non
\end{eqnarray}
where $N={p\choose n}$ is the nucleon doublet of two-component
spinors, the subscripts $d$ and $t$ denote the ${}^3\mathrm{S}_1$ and ${}^1\mathrm{S}_0$
channel of $NN$ scattering, and $M$ is the average nucleon
mass. $P^i_d$ and $P^A_t$ are the projectors onto the $^3 \mathrm{S}_1$ and $^1
\mathrm{S}_0$ channels, respectively
\begin{equation}\label{proj}
  \left(P^i_d\right)^{b\beta}_{a\alpha}=
  \frac{1}{\sqrt{8}}\; (\sigma_2\sigma^i)_\alpha^\beta
  \;(\tau_2)_a^b \;\;,\;\;
  \left(P^A_t\right)^{b\beta}_{a\alpha}=
  \frac{1}{\sqrt{8}}\; (\sigma_2)_\alpha^\beta
  \;(\tau_2\tau^A)_a^b \;\;,
\end{equation}
with $\sigma$ ($\tau$) the Pauli matrices acting in spin (isospin)
space. The coefficients $y_i$,
$\Delta^{(-1)}_i$ and $\Delta^{(0)}_i$ encode all short distance
physics -- like pion and $\omega$ exchanges, quarks and gluons, and
resonances like the $\Delta$. Here it is necessary to split both
$\Delta$'s into leading ($\Delta^{(-1)}$) and subleading pieces
($\Delta^{(0)}$).

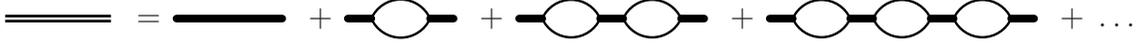
\begin{figure}[!htb]
  \begin{center}
    \feynbox{40\unitlength}{
            \begin{fmfgraph*}(40,40)
              \fmfleft{i} \fmfright{o} \fmf{double,width=thin}{i,o}
            \end{fmfgraph*}}
          \hq$=$\hq \feynbox{40\unitlength}{
            \begin{fmfgraph*}(40,40)
              \fmfleft{i} \fmfright{o} \fmf{vanilla,width=1.5*thick}{i,o}
            \end{fmfgraph*}}
          \hq$+$\hq \feynbox{40\unitlength}{
            \begin{fmfgraph*}(40,40)
              \fmfleft{i} \fmfright{o}
              \fmf{vanilla,width=1.5*thick,tension=5}{i,v1}
              \fmf{vanilla,width=1.5*thick,tension=5}{o,v2}
              \fmf{vanilla,width=thin,left=0.65}{v1,v2}
              \fmf{vanilla,width=thin,left=0.65}{v2,v1}
            \end{fmfgraph*}}
          \hq$+$\hq \feynbox{70\unitlength}{
            \begin{fmfgraph*}(70,40)
              \fmfleft{i} \fmfright{o}
              \fmf{vanilla,width=1.5*thick,tension=5}{i,v1}
              \fmf{vanilla,width=1.5*thick,tension=5}{v2,v3}
              \fmf{vanilla,width=1.5*thick,tension=5}{v4,o}
              \fmf{vanilla,width=thin,left=0.65}{v1,v2}
              \fmf{vanilla,width=thin,left=0.65}{v2,v1}
              \fmf{vanilla,width=thin,left=0.65}{v3,v4}
              \fmf{vanilla,width=thin,left=0.65}{v4,v3}
            \end{fmfgraph*}}
          \hq$+$\hq \feynbox{100\unitlength}{
            \begin{fmfgraph*}(100,40)
              \fmfleft{i} \fmfright{o}
              \fmf{vanilla,width=1.5*thick,tension=5}{i,v1}
              \fmf{vanilla,width=1.5*thick,tension=5}{v2,v3}
              \fmf{vanilla,width=1.5*thick,tension=5}{v4,v5}
              \fmf{vanilla,width=1.5*thick,tension=5}{v6,o}
              \fmf{vanilla,width=thin,left=0.65}{v1,v2}
              \fmf{vanilla,width=thin,left=0.65}{v2,v1}
              \fmf{vanilla,width=thin,left=0.65}{v3,v4}
              \fmf{vanilla,width=thin,left=0.65}{v4,v3}
              \fmf{vanilla,width=thin,left=0.65}{v5,v6}
              \fmf{vanilla,width=thin,left=0.65}{v6,v5}
            \end{fmfgraph*}}
          \hq$+\;\dots$
  \end{center}
  \caption{\label{fig:deuteronpropagator} \sl The deuteron propagator
    at LO from the Lagrangian (\protect\ref{dlag}). The thick solid
    line denotes the bare propagator \protect$\frac{-\ii}{\Delta^{(-1)}_d}$,
    the double line its dressed counterpart.}
\end{figure}

Since the theory is nonrelativistic, all particles propagate forward in time,
the nucleon tadpoles vanish, and the propagator for the nucleon fields is
\beq
\label{nucprop}
iS(p)=\frac{i}{p_0-p^2/2M +i\epsilon}\,.
\eeq
The deuteron and dibaryon propagators are more complicated because of the coupling
to two-nucleon states. For instance, the bare deuteron propagator is simply a constant, 
$-i/\Delta_d^{-1}$, but the full propagator gets dressed by nucleon loops
to all orders as illustrated in Fig.~\ref{fig:deuteronpropagator}.

The nucleon-loop integral has a linear 
ultraviolet (UV) divergence which can be absorbed 
in $y^2_d/\Delta^{-1}_d$, and a finite piece determined by the unitarity 
cut. Summing the resulting geometric series leads to the deuteron (and
analogously the dibaryon) propagators in terms
of physical quantities at LO:
\begin{eqnarray}\label{props}
  \ii\triangle^{ij}_{d}(p) \;=\;\ii\delta^{ij}\,\triangle_{d}(p)& =&
  \frac{4\pi \ii}{M y^2_{d}}\;
  \frac{\delta^{ij}}{\gamma_{d}-\sqrt{\frac{\pv^2}{4}-M
  p_0-\ii\varepsilon}}\;\;,\non\\
  \ii\triangle^{AB}_{t}(p) \;=\;\ii\delta^{AB}\,\triangle_{t}(p)& =&
  \frac{4\pi \ii}{M y^2_{t}}\;
  \frac{\delta^{AB}}{\gamma_{t}-\sqrt{\frac{\pv^2}{4}-M
  p_0-\ii\varepsilon}}\;\;.
\end{eqnarray}
$\gamma_d=\sqrt{M B} = 45.7066\;\MeV$ and the deuteron binding energy
is $B=2.225\;\MeV$. The typical momentum $\gamma_t=1/a_t$ of the
virtual bound state is extracted from the scattering length in this
channel, $a_t=-23.714\;\fm$.

\section{\mbox{\boldmath $S$ = $1/2$ $nd$} Scattering}

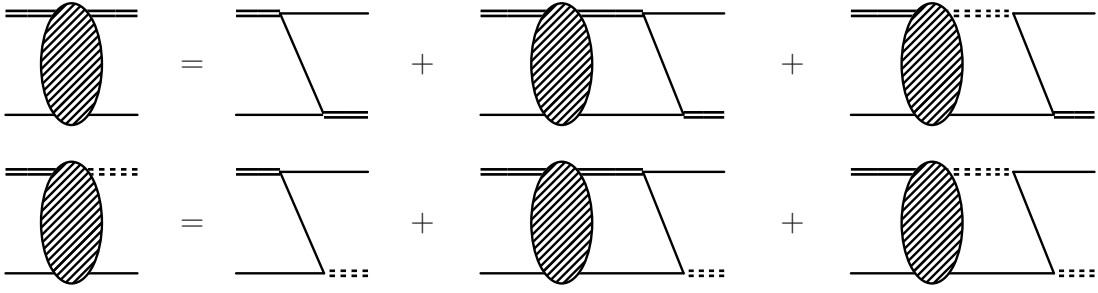
\begin{figure}[!htb]
  \begin{center}
    
    \vspace*{3ex}
    
    \setlength{\unitlength}{0.6pt}
    
    \feynbox{104\unitlength}{
            \begin{fmfgraph*}(104,64)
              \fmfleft{i2,i1} \fmfright{o2,o1}
              \fmf{double,width=thin,tension=3}{i1,v1}
              \fmf{double,width=thin,tension=1.5}{v1,v3,v2}
              \fmf{double,width=thin,tension=3}{v2,o1}
              \fmf{vanilla,width=thin}{i2,v4,o2} \fmffreeze
              \fmf{ellipse,rubout=1,label=$\fs t_{\mathrm{d}}^{ij}$,
                label.dist=0.25w,label.side=right}{v3,v4}
              \end{fmfgraph*}}
            \hq$=$\hq \feynbox{104\unitlength}{
            \begin{fmfgraph*}(104,64)
              \fmfleft{i2,i1} \fmfright{o2,o1}
              \fmf{double,width=thin,tension=4}{i1,v1,v2}
              \fmf{vanilla,width=thin}{v2,o1}
              \fmf{double,width=thin,tension=4}{v3,v4,o2}
              \fmf{vanilla,width=thin}{i2,v3} \fmffreeze
              \fmf{vanilla,width=thin}{v2,v3}
              \end{fmfgraph*}}
            \hq$+$\hq \feynbox{192\unitlength}{
            \begin{fmfgraph*}(192,64)
              \fmfleft{i2,i1} \fmfright{o2,o1}
              \fmf{double,width=thin,tension=3}{i1,v1}
              \fmf{double,width=thin,tension=1.5}{v1,v6,v5}
              \fmf{double,width=thin,tension=3}{v5,v2}
              \fmf{vanilla,width=thin}{v2,o1} \fmf{vanilla,width=thin}{i2,v7}
              \fmf{vanilla,width=thin,tension=0.666}{v7,v4}
              \fmf{double,width=thin,tension=4}{v4,v3,o2} \fmffreeze
              \fmf{vanilla,width=thin}{v4,v2} \fmf{ellipse,rubout=1,label=$\fs
                t_{\mathrm{d}}^{ij}$,
                label.dist=0.15w,label.side=right}{v6,v7}
              \end{fmfgraph*}}
            \hq$+$\hq \feynbox{192\unitlength}{
            \begin{fmfgraph*}(192,64)
              \fmfleft{i2,i1} \fmfright{o2,o1}
              \fmf{double,width=thin,tension=3}{i1,v1}
              \fmf{double,width=thin,tension=1.5}{v1,v6}
              \fmf{dbl_dashes,width=thin,tension=1.5}{v6,v5}
              \fmf{dbl_dashes,width=thin,tension=3}{v5,v2}
              \fmf{vanilla,width=thin}{v2,o1} \fmf{vanilla,width=thin}{i2,v7}
              \fmf{vanilla,width=thin,tension=0.666}{v7,v4}
              \fmf{double,width=thin,tension=4}{v4,v3,o2} \fmffreeze
              \fmf{vanilla,width=thin}{v4,v2} \fmf{ellipse,rubout=1,label=$\fs
                t_{\mathrm{t}}^{iA}$,
                label.dist=0.15w,label.side=right}{v6,v7}
              \end{fmfgraph*}}
            
            \vspace*{4ex}
            
            \feynbox{104\unitlength}{
            \begin{fmfgraph*}(104,64)
              \fmfleft{i2,i1} \fmfright{o2,o1}
              \fmf{double,width=thin,tension=3}{i1,v1}
              \fmf{double,width=thin,tension=1.5}{v1,v3}
              \fmf{dbl_dashes,width=thin,tension=1.5}{v2,v3}
              \fmf{dbl_dashes,width=thin,tension=3}{o1,v2}
              \fmf{vanilla,width=thin}{i2,v4,o2} \fmffreeze
              \fmf{ellipse,rubout=1,label=$\fs t_{\mathrm{t}}^{iA}$,
                label.dist=0.25w,label.side=right}{v3,v4}
              \end{fmfgraph*}}
            \hq$=$\hq \feynbox{104\unitlength}{
            \begin{fmfgraph*}(104,64)
              \fmfleft{i2,i1} \fmfright{o2,o1}
              \fmf{double,width=thin,tension=4}{i1,v1,v2}
              \fmf{vanilla,width=thin}{v2,o1}
              \fmf{dbl_dashes,width=thin,tension=4}{o2,v4,v3}
              \fmf{vanilla,width=thin}{i2,v3} \fmffreeze
              \fmf{vanilla,width=thin}{v2,v3}
              \end{fmfgraph*}}
            \hq$+$\hq \feynbox{192\unitlength}{
            \begin{fmfgraph*}(192,64)
              \fmfleft{i2,i1} \fmfright{o2,o1}
              \fmf{double,width=thin,tension=3}{i1,v1}
              \fmf{double,width=thin,tension=1.5}{v1,v6,v5}
              \fmf{double,width=thin,tension=3}{v5,v2}
              \fmf{vanilla,width=thin}{v2,o1} \fmf{vanilla,width=thin}{i2,v7}
              \fmf{vanilla,width=thin,tension=0.666}{v7,v4}
              \fmf{dbl_dashes,width=thin,tension=4}{o2,v3,v4} \fmffreeze
              \fmf{vanilla,width=thin}{v4,v2} \fmf{ellipse,rubout=1,label=$\fs
                t_{\mathrm{d}}^{ij}$,
                label.dist=0.15w,label.side=right}{v6,v7}
              \end{fmfgraph*}}
            \hq$+$\hq \feynbox{192\unitlength}{
            \begin{fmfgraph*}(192,64)
              \fmfleft{i2,i1} \fmfright{o2,o1}
              \fmf{double,width=thin,tension=3}{i1,v1}
              \fmf{double,width=thin,tension=1.5}{v1,v6}
              \fmf{dbl_dashes,width=thin,tension=1.5}{v6,v5}
              \fmf{dbl_dashes,width=thin,tension=3}{v5,v2}
              \fmf{vanilla,width=thin}{v2,o1} \fmf{vanilla,width=thin}{i2,v7}
              \fmf{vanilla,width=thin,tension=0.666}{v7,v4}
              \fmf{dbl_dashes,width=thin,tension=4}{o2,v3,v4} \fmffreeze
              \fmf{vanilla,width=thin}{v4,v2} \fmf{ellipse,rubout=1,label=$\fs
                t_{\mathrm{t}}^{iA}$,
                label.dist=0.15w,label.side=right}{v6,v7}
              \end{fmfgraph*}}
            
            \vspace*{1ex}
            
            \setlength{\unitlength}{1pt}
    
  \end{center}
    \caption{\label{fig:LOfaddeevdoublet} \sl The coupled set of Faddeev
      equations
      which need to be solved for \protect$t_{\mathrm{d},\mathrm{t}}$ at LO in the
      doublet channel. The double dashed line denotes the dibaryon field
      \protect$t^{iA}_t$.}
\end{figure}

The Lagrangian (\ref{dlag}) can now be used to describe the
$nd$-scattering in the $S=1/2$ channel. The treatment involves coupled
channel equations both in the $^3\mathrm{S}_1$ and $^1\mathrm{S}_0$
partial waves. The spin zero dibaryon field $t$ also contributes in
intermediate states of $nd$ amplitudes. It is possible to obtain a
system of two coupled integral equations (previously derived 
using a different method \cite{skorny}) for the
$d+N\rightarrow d+N$ amplitude $i t^{ij}_d(\vec{k},\vec{p},\epsilon)$
and for the $d+N\rightarrow t+N$ amplitude $i
t^{iA}_t(\vec{k},\vec{p},\epsilon)$, pictorially represented in
Fig.~\ref{fig:LOfaddeevdoublet}.

A momentum cutoff $\Lambda$ has to be introduced in the integral
equations. In the limit $\Lambda\to \infty$ these equations do not
have a unique solution because the phase of the asymptotic solution is
undetermined \cite{danilov}. For a finite $\Lambda$ this phase is
fixed and the solution is unique. However, the equations with a cutoff
display a strong cutoff dependence that does not appear in any order
in perturbation theory. The amplitude $t_d(p,k=\mbox{const.})$ shows a
strongly oscillating behavior \cite{3stooges}. This cutoff dependence
is not created by divergent Feynman diagrams. It is a nonperturbative
effect and appears although all individual diagrams are superficially
UV finite. For an extended discussion on this effect, see
\cite{3stooges}.

The solution is to add one-parameter 
three-body force counterterm $H(\Lambda)/\Lambda^2$ 
that runs with the cutoff $\Lambda$ \cite{bosons}. This counterterm
represents a three-body force which is obtained by including
\begin{eqnarray}
\label{3bod}
{\cal L}_3 = -\frac{2MH(\Lambda)}{\Lambda^2}  \! \! \! \! \! \! \! \! &&
 \left\{ \frac{1}{2}{y_d^2} N^\dagger 
({d^i}{\sigma}_i)^\dagger ({d^i}{\sigma}_i) N \right. \non \\
&& +\frac{1}{6} y_d y_t \left[ N^\dagger ({d^i}{\sigma}_i)^\dagger
(t^A \tau_A) N + h.c. \right]  \\
&& \left. +\frac{1}{2}{y_t^2} N^\dagger
(t^A \tau_A)^\dagger (t^A \tau_A) N \right\}\, \non
\end{eqnarray}
in the Lagrangian (\ref{dlag}). This yields the equations:
\begin{eqnarray}\label{inteh1}
  t_d(k,p)&=&\;4 y^2_d M
\left[K(p,k)+\frac{2H(\Lambda)}{\Lambda^2}\right]+
\frac{1}{\pi}\int\limits_0^\Lambda dq\; q^2\;
  t_d(k,q)\;
  \frac{1}{\sqrt{\frac{3 q^2}{4}-ME-
     \ii\varepsilon}-\gamma_d}\;\times\nonumber\\
&&\times\; \left[K(p,q)+\frac{2H(\Lambda)}{\Lambda^2}\right]-
\;\frac{3}{\pi}\frac{y_d}{y_t}\int\limits_0^\Lambda dq\; q^2\;
 t_t(k,q)\;
 \frac{1}{\sqrt{\frac{3 q^2}{4}-ME- \ii\varepsilon}-\gamma_t}\;
 \times\nonumber\\
 &&\times\; \left[K(p,q)+\frac{2H(\Lambda)}{3\Lambda^2}\right]\;\;,
\end{eqnarray}
\begin{eqnarray}\label{inteh2}
  t_t(k,p)&=&-\;12 {y_d y_t M}
  \left[K(p,k)+\frac{2H(\Lambda)}{3\Lambda^2}\right]+\;\frac{1}{\pi}\int\limits_0^\Lambda dq\; q^2\;
 t_t(k,q)\;\frac{1}{\sqrt{\frac{3 q^2}{4}-ME- \ii\varepsilon}-\gamma_t}\;
 \times \nonumber\\
 &&\times\; \left[K(p,q)+\frac{2H(\Lambda)}{\Lambda^2}\right]-
 \;\frac{3}{\pi}\frac{y_t}{y_d}\int\limits_0^\Lambda dq\; q^2\;
 t_d(k,q)\;
 \frac{1}{\sqrt{\frac{3 q^2}{4}-ME- \ii\varepsilon}-\gamma_d}
 \;\times\nonumber\\
 &&\times\;\;
 \left[K(p,q)+\frac{2H(\Lambda)}{3\Lambda^2}\right]\;\;,
\end{eqnarray}
where $k$ ($p$) denote the incoming (outgoing) momenta in the center-of-mass 
frame, $M E = 3k^2/4 - \gamma_d^2$ is the total energy, 
the kernel $K(p,q)$ is given by
\beq
K(p,q)=\frac{1}{2pq}\ln\left(\frac{q^2+pq-p^2-ME}{q^2-pq-p^2-ME}\right)\,,
\eeq
and $H(\Lambda)$ is defined as follows:
\beq
\label{runH}
H(\Lambda)=-    \frac{\sin[s_0\ln({\Lambda}/{\Lambda_\star})-
                   {\rm arctg}(1/s_0)]}
                 {\sin[s_0 \ln({\Lambda}/{\Lambda_\star})+
                   {\rm arctg}(1/s_0)]} \;.
\eeq
In eq. (\ref{runH}) $s_0\approx 1.0064$ 
and $\Lambda_*$ is a dimensionful parameter that determines 
the asymptotic phase of the off-shell amplitude
\cite{bosons}, and is fitted to reproduce the experimental value
for the $S=1/2$ $nd$ scattering 
length, $a_3^{(1/2)}=(0.65 \pm 0.04)\mbox{ fm}$ \cite{dilg},
yielding $\Lambda_* = 0.9 \mbox{ fm}^{-1}$.

\section{Effective Range Expansion}

The cutoff dependence is expected to be eliminated from the equations
if the effective range parameters $\rho_d$ and $r_{0t}$ are included
in the effective range expansion and are treated like the scattering
length $a$, \ie\ taken of order $Q^{-1}$, and summed to all orders.
Ref. \cite{besa} gave reasons to conclude that this power counting
should improve the overall convergence. In the ${}^3\mathrm{S}_1$
channel, the expansion around the deuteron pole gives
\begin{equation}
   k \cot(\delta)=-\gamma_d+\frac{1}{2} \rho_d ( k^2+\gamma^2_d) + \dots
\end{equation}
for the nucleon-nucleon phase shifts. In the singlet channel
${}^1\mathrm{S}_0$, no real bound state exists, so the condition to
impose is that the effective range expansion
\begin{equation}
  \label{acondition}
   k \cot(\delta)=-{1 \over a_t}+\frac{1}{2} r_{0t} k^2 + \dots
\end{equation}
is satisfied. Resumming the effective ranges $\rho_d$ and $r_{0t}$ in the
full deuteron and dibaryon propagators generates
\begin{eqnarray}
\label{full}
  \ii\triangle^{\rho_d,ij}_{d}(p) \;=\;\ii\delta^{ij}\,\triangle^{(\rho_d)}_{d}(p)
  &=&-\;\frac{4\pi \ii}{M y^2_d}\;
  \frac{\delta^{ij}}{-\gamma_{d}+\frac{1}{2} \rho_d (M p_0-\frac{\pv^2}{4}
  +\gamma^2_d)+
  \sqrt{\frac{\pv^2}{4}-M p_0-\ii\varepsilon}}\;\;,\non\\
  \smallskip
  \ii\triangle^{r_{0t},AB}_{t}(p) \;=\;\ii\delta^{AB}\,\triangle^{(r_{0t})}_{t}(p)
  &=&-\;\frac{4\pi \ii}{M y^2_t}\;
  \frac{\delta^{AB}}{-\gamma_t+\frac{1}{2} r_{0t} (M p_0-\frac{\pv^2}{4})+
  \sqrt{\frac{\pv^2}{4}-M p_0-\ii\varepsilon}}\;\;. \non \\
\end{eqnarray}
The numerical values $\rho_d=1.765\;\fm$ and $r_{0t}=2.73\;\fm$
\cite{nij} have been used\footnote{For an early computation including
effective ranges and comparisons with potential models, see \cite{ss}.}.

It is still possible to introduce in the computation a three-body
term, which now is independent from the cutoff $\Lambda$ and is in
fact a constant contact interaction. The resulting integral equations
are completely analogous to (\ref{inteh1}), (\ref{inteh2}), but now
with the propagators (\ref{full}). The integration is carried on the
full real positive axis without any cutoff. The integral equations are
then solved numerically using the techniques outlined in \cite{pbhg}.

\begin{figure}[htb]
\centerline{
\begin{minipage}[t]{.44\linewidth}\centering
\mbox{\epsfig{file=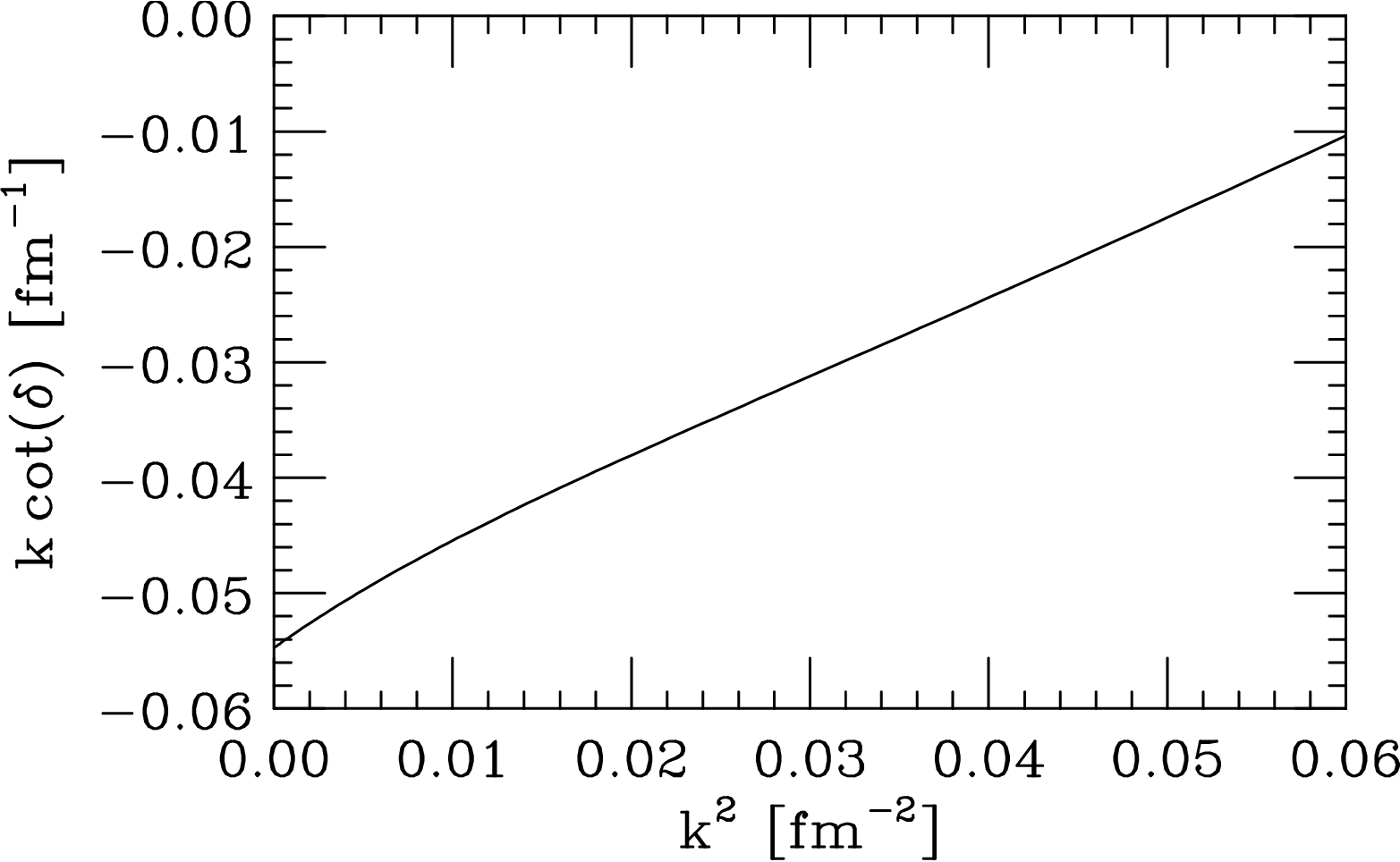,width=7.25cm}}
\caption{Energy dependence for $S=1/2$ $nd$ scattering with $H = 0$ or
with $H \neq 0$ and $\Lambda \rightarrow \infty$ when the effective
ranges are included in the calculation.}
\label{fig:badres}
\end{minipage}
\hspace{0.6cm}
\begin{minipage}[t]{.44\linewidth}\centering
\mbox{\epsfig{file=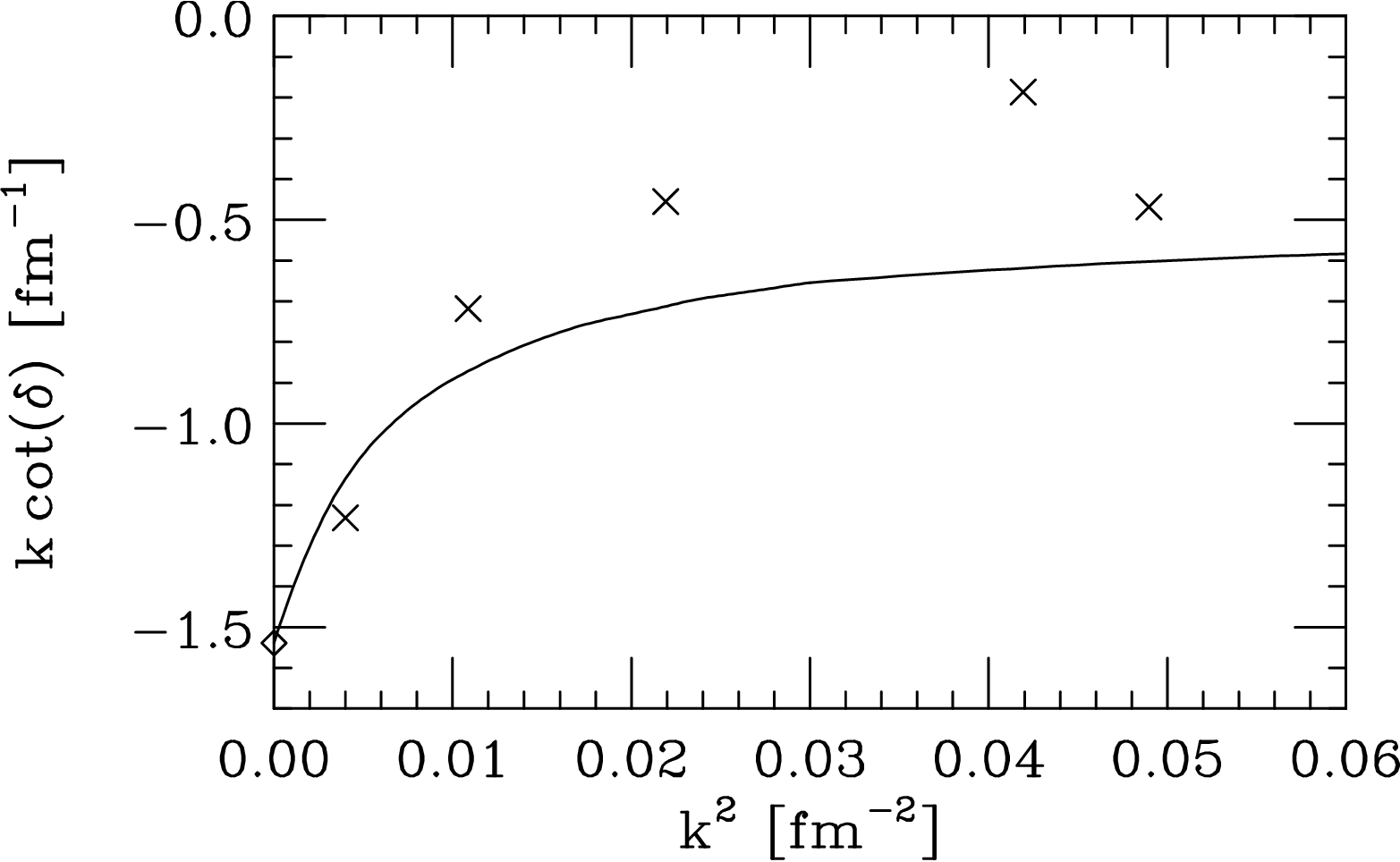,width=7.25cm}}
\caption{Cutoff-insensitive energy dependence for $S=1/2$ $nd$
scattering for $\Lambda_*=0.9\mbox{ fm}^{-1}$, as given in
ref. \protect\cite{3stooges2}, without resumming the effective ranges.
Data are from the phase-shift analysis of van Oers and Seagrave
\protect\cite{vOers} (crosses) and a measurement by Dilg {\it et al.}
\protect\cite{dilg} (diamond).}
\label{fig:bhvk}
\end{minipage}}
\end{figure}

The solutions for the amplitudes coming from the equations do not show any
visible dependence on the three-body interaction term $H$. It is possible to
extract the energy dependence for $k \cot(\delta)$, where $k$ is the
incoming momentum of the particles in the center of mass frame, from
the behavior of the on-shell amplitude $t_d(k,k)$. Results are plotted
on Fig. \ref{fig:badres}. Compare with the results given in
ref. \cite{3stooges2} (Fig. \ref{fig:bhvk}).

The explanation for this outcome stems from the fact that, while the
convergence of the equations has improved because of the inclusion of
the effective ranges (which multiply a factor $\propto$ $q^2$ in the
denominator of the propagators), this is not yet sufficient to make
the equations completely convergent. After subtracting from the
amplitudes $t_{d,t}(k,p)$ the parts satisfying the (convergent)
equations {\it without} the three-body term $H$, the remaining parts
$t^{\prime}_{d,t}(k,p)$ are described by equations containing linear
divergences (plus convergent terms). Regularizing these divergences by
introducing a cutoff $\Lambda$ drives $t^{\prime}_{d,t}(k,p)$ to
approximate quantities $\propto$ $1/\Lambda$, \ie\ to negligibly small
numbers if the cutoff is set sufficiently
high. $t_{d,t}(k,p)$ are then effectively cutoff-independent,
but independent from $H$ as well.

\section{Conclusions}

The study of three-nucleon systems using EFT methods in the $S=1/2$
channel is more complicated than for the $S=3/2$ channel. While for
the latter $nd$ scattering accurate predictions are obtained
\cite{3stooges2}, the $S=1/2$ channel displays a strong cutoff
dependence even though all individual diagrams are UV finite. This
dependence can be eliminated at LO only for the equations without
three-body interactions, if the effective range parameters, obtained
from the effective range expansion in $N N$ scattering, are taken into
account in the deuteron and dibaryon propagators. This is achieved
modifying the usual power-counting scheme, assuming $\rho_d$ and $r_{0t}$
of order $Q^{-1}$ like the scattering length $a$.

Unfortunately this improvement is not sufficient to eliminate all the
infinities originated in the loop diagrams from the integral
equations. The remaining divergent terms have the effect of driving
the three-body-dependent parts of the amplitudes to negligible values,
while the amplitude parts dependent only on the finite equations
(those {\it without} the three-body term) prevail.

This makes eqs. (\ref{inteh1}), (\ref{inteh2}), after the inclusion of
the effective range parameters, insensitive to the three-body term
necessary to describe the influence of the triton bound state and to
reproduce the experimental data for $nd$ scattering in the S-wave
doublet channel.

Alternatively, one can argue that the three-body force term arise not
at LO but at NLO. Yet in this case it must generate a contribution to
the amplitude much bigger than the $\sim$~30~\% variation one
expects going from LO to NLO.

Therefore this test of the new power-counting
procedure does not yield the successful results achieved in the
two-body problem case. Yet it is possible that a better treatment of
this problem implies the explicit inclusion of the triton propagator
in the theory, analogously to what has been done so far for the
deuteron and the dibaryon, and according to the spirit of
ref. \cite{besa}.

\section*{Acknowledgments}
I would like to thank Paulo Bedaque for several discussions and Silas
Beane for comments. This research was supported in part by the
U.S. Department of Energy grant DE-FG02-01ER41155.


\end{fmffile}
\end{document}